\documentclass{article}

\usepackage{arxiv}

\usepackage{bm}
\usepackage{subcaption}
\usepackage{amsmath}
\usepackage{amsfonts}
\usepackage{listings}

\usepackage[svgnames]{xcolor}

\lstdefinelanguage{XML}
{
  basicstyle=\ttfamily\footnotesize,
  morestring=[b]",
  moredelim=[s][\bfseries\color{Maroon}]{<}{\ },
  moredelim=[s][\bfseries\color{Maroon}]{</}{>},
  moredelim=[l][\bfseries\color{Maroon}]{/>},
  moredelim=[l][\bfseries\color{Maroon}]{>},
  morecomment=[s]{<?}{?>},
  morecomment=[s]{<!--}{-->},
  commentstyle=\color{DarkOliveGreen},
  stringstyle=\color{blue},
  identifierstyle=\color{red}
}

\usepackage{color}

\usepackage{pgfplots}
\usepackage{pgfplotstable}
\pgfplotsset{
compat=newest, 
tick label style={font=\footnotesize}, 
}

\usepackage{overpic}

\usepackage{tikz}
\usetikzlibrary{mindmap,shadows}
\usetikzlibrary{arrows,intersections}
\usepackage{smartdiagram}

\usepackage{booktabs}
\usepackage[]{units}

\title{Implementation of an aeroacoustic simulation pipeline using openCFS-Acoustics and openCFS-Data applied to human phonation}

\author{ Stefan Schoder \\
	Group of Aeroacoustics and Vibroacoustics, IGTE\\
	TU Graz\\
	Inffeldgasse 18, 8010 Graz \\
	\texttt{stefan.schoder@tugraz.at} \\
}

\renewcommand{\shorttitle}{openCFS Software}

\usepackage{hyperref}
\hypersetup{
pdftitle={openCFS Software},
pdfsubject={preprint},
pdfauthor={Schoder, Roppert},
pdfkeywords={FEM Sotware, open Source},
}

\begin{document}
\maketitle

\begin{abstract}
The human phonation process be modeled using the Finite Element Method (FEM) which provides a detailed representation of the voice production process. A software implementation in C++ using FEM (openCFS) has been used to simulate the phonation process. The FEM model consists of a 3D mesh of the upper human airways. The simVoice model provides an accurate representation of the phonation process and was valid in several publications. In this article, we show how to set up the model using openCFS and openCFS-Data.
\end{abstract}

\keywords{Open Source FEM Software \and Multiphysics Simulation \and C++ \and Acoustics \and Aero-Acoustics \and openCFS \and Human Phonation \and Voice}

%

\section{Introduction}
\label{sec:Intro} 

The human phonation process, which involves the production of speech sounds through the vibration of the vocal folds in the larynx, is a complex and dynamic phenomenon that has been the subject of extensive research. Understanding the underlying dynamics of this process is essential for developing technologies that can improve speech synthesis and recognition, as well as diagnose and treat speech disorders.

One approach to modeling the phonation process is through an acoustic model, which describes the sound produced by the vocal folds in response to given flow data. The acoustic model is based on the Finite Element Method (FEM). Within this contribution, we concentrate on the openCFS-Acoustic \cite{CFS} module \textit{openCFS-Data} \cite{CFSDAT}. Regarding the hybrid aeroacoustic modeling approach in general, it was applied to and validated for cavity noise simulations \cite{schoder2018aeroacoustic,schoder2019hybrid,schoder2020conservative}, mixing layer noise \cite{schoder2022aeroacoustic,schoder2022cpcwe}, automotive applications \cite{engelmann2020generic,freidhager2021simulationen,schoder2020numerical,weitz2019numerical,maurerlehner2022aeroacoustic}. Furthermore, the hybrid aeroacoustic workflow was found to be useful for fan noise computations \cite{schoder2020computational,tautz2018source,schoder2021application,tieghi2022machine,tieghi2023machine,schoder2022dataset} and the noise emissions of the turbocharger compressor \cite{freidhager2022applicability,kaltenbacher2020modelling,freidhager2020influences}. In particular, the acoustics of fluid-structure-acoustic-interaction processes of human phonation was simulated and validated, leading to simVoice \cite{schoder2020hybrid,valavsek2019application,zorner2016flow,schoder2021aeroacoustic,falk20213d,lasota2021impact,maurerlehner2021efficient,schoder2022learning,lasota2023anisotropic,schoder2022pcwe,kraxberger2022machine}. Based on this rich expertise on simulating the human voice, we show how to develop such a model using openCFS.

\section{Human Phonation Simulation}
In this example, we will simulate the acoustic wave propagation inside a 3D model of the human vocal tract. This example is based on the simulation workflow that was used for \textbf{simVoice}.
Thereby, the acoustic sources for the used PCWE are computed from the incompressible pressure field that was obtained by a CFD simulation of the flow. 

\subsection{Mesh}
\begin{figure}
    \centering
    \includegraphics[scale=0.4]{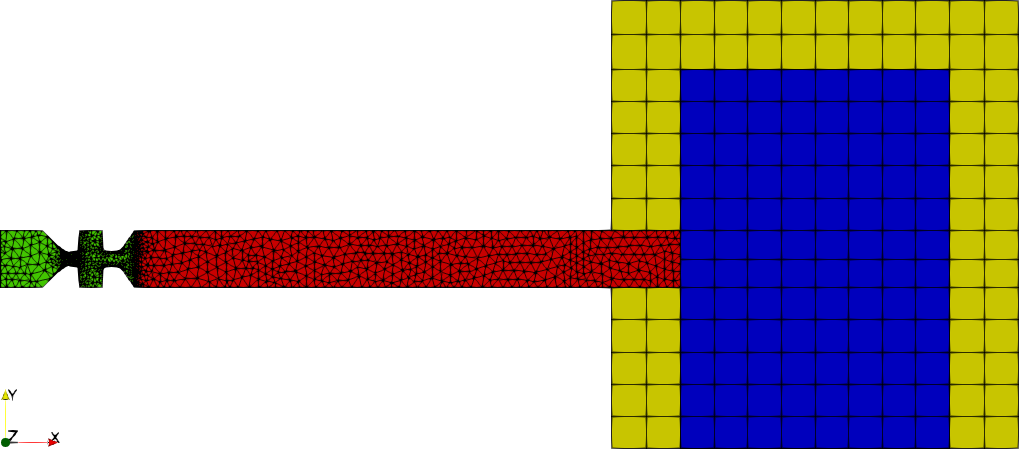}
    \caption{The mesh of the validation setup.}
    \label{fig:valid}
\end{figure}

The acoustic Mesh was created with the commercial software \textit{Ansys ICEM}. It consists of 4 regions (see Fig. \ref{fig:valid}): The larynx region LARYNX, the vocal tract VT, the propagation region PR, and the PML.

\subsection{Pressure Interpolation}
In order to perform an acoustic simulation on the presented mesh, the acoustic source data has to be interpolated onto it. Therefore, openCFS-Data (a package included in openCFS) is used. As openCFS does, openCFS-Data uses an XML file \footnote{https://opencfs.gitlab.io/userdocu/Applications/Singlefield/Acoustics/HumanPhonation/interpolatePressure.xml} to configure the workflow. The basic structure including XML header reads the following:
\begin{lstlisting}[language=XML]
<?xml version="1.0" encoding="UTF-8"?>
<cfsdat xmlns:xsi="http://www.w3.org/2001/XMLSchema-instance" xmlns="http://www.cfs++.org/simulation">
	<pipeline>
	
		USERINPUT
	
	</pipeline>
</cfsdat>
\end{lstlisting} 

First, we define the length and time step size of the CFD data. We simulate one full cycle of the vocal folds' motion, which are oscillation at $f_0=148 \mathrm{Hz}$. Therefore, we define \textbf{numsteps}$= 675$ time steps with a step size of $\Delta t=1\cdot 10^{-5} s$.
\begin{lstlisting}[language=XML]
<stepValueDefinition>
	<startStop>
		<startStep value="0"/>              
  <!-- same as the first Ensight step number, but ATTENTION consider delta t !!! -->
		<numSteps value="675"/>           
  <!-- step numbers to calculate -->
		<startTime value="1e-05"/>          
  <!-- start from the time that is equal to the 7501th step from the Ensight -->
		<delta value="1e-05"/>              
  <!-- delta t -->
		<deleteOffset value="no"/>          
  <!-- if NO then the time will be the same as in the CFD (Ensight) -->
	</startStop>
</stepValueDefinition>
\end{lstlisting} 

The CFD data was exported in \textit{Ensight-Gold} format as the file: 'CFD\_results.case', which can be obtained by any CFD software.
\begin{lstlisting}[language=XML]
<meshInput id="input" gridType="fullGrid">
	<inputFile>
		<ensight fileName="cfd_data_path/CFD_results.case">
			<variableList>
				<variable CFSVarName="fluidMechPressure" EnsightVarName="Pressure"/>
			</variableList>
		</ensight>
	</inputFile>
</meshInput>
\end{lstlisting} 


For the field interpolation, the conservative cut-cell volume approach is used to interpolate the cell data onto the acoustic mesh, saved in the hdf5 file 'CAA\_mesh.h5'.
\begin{lstlisting}[language=XML]
<interpolation type="FieldInterpolation_Conservative_CutCell" id="interp" inputFilterIds="input">
	<targetMesh>
		<hdf5 fileName="CAA_mesh.h5"/>
	</targetMesh>
	<singleResult>
		<inputQuantity resultName="fluidMechPressure"/>
		<outputQuantity resultName="fluidMechPressure_interpolated"/>
	</singleResult>
	<regions>
		<sourceRegions>
			<region name="Background"/>
		</sourceRegions>
		<targetRegions>
			<region name="LARYNX"/>
			<region name="VT"/>
		</targetRegions>
	</regions>
</interpolation>
\end{lstlisting} 


Finally, we define the output filename by \textbf{meshOutput-id} and the interpolated pressure field for both regions, the LARYNX and the VT (vocal tract).
\begin{lstlisting}[language=XML]
<meshOutput id="InterpolatedPressureField" inputFilterIds="interp">
	<outputFile>
		<hdf5 extension="cfs" compressionLevel="6" externalFiles="no"/> 
  <!-- compression level (greater number = bigger compression; default=1) -->
	</outputFile>
	<saveResults>
		<result resultName="fluidMechPressure_interpolated">
			<allRegions/>
		</result>
	</saveResults>
</meshOutput>
\end{lstlisting} 

Finally, run the openCFS-Data interpolation inside the shell by the command \textit{cfsdat interpolatePressure}.

\subsection{Source term computation}
As we now have obtained the conservatively interpolated incompressible flow pressure on the acoustic grid, we again use openCFS-Data to compute the RHS source of the PCWE $\partial p / \partial t$. The convective part of the RHS source is neglected in this example due to its negligible impact on the resulting sound spectrum in the far field (see \cite{schoder2021aeroacoustic}). The second XML file\footnote{https://opencfs.gitlab.io/userdocu/Applications/Singlefield/Acoustics/HumanPhonation/calc\_dpdt.xml} is structured as before with the inputs:

We define the length and time step size of the CFD data, as done for the flow pressure Interpolation.
\begin{lstlisting}[language=XML]
<stepValueDefinition>
	<startStop>
		<startStep value="0"/>              
  <!-- same as the first Ensight step number -->
		<numSteps value="675"/>             
  <!-- step numbers to calculate -->
		<startTime value="1e-05"/>          
  <!-- start from the time that is equal to the *startStep*-th step from the Ensight -->
		<delta value="1e-05"/>              
  <!-- delta t -->
		<deleteOffset value="no"/>          
  <!-- if NO then the time will be the same as in the CFD (Ensight) -->
	</startStop>
</stepValueDefinition>
\end{lstlisting} 

This time, we define the previously computed interpolated flow pressure data.
\begin{lstlisting}[language=XML]
<meshInput id="input" gridType="fullGrid">
	<inputFile>
		<hdf5 fileName="results_hdf5/InterpolatedPressureField.cfs"/>
	</inputFile>
</meshInput>
\end{lstlisting} 

Then, define the time derivative filter. For more information, please have a look at \cite{CFSDAT}.
\begin{lstlisting}[language=XML]
<timeDeriv1 inputFilterIds="input" id="pressure_time_derivative">
	<singleResult>
		<inputQuantity resultName="fluidMechPressure_interpolated"/>
		<outputQuantity resultName="acouRhsLoad"/>
	</singleResult>
</timeDeriv1>
\end{lstlisting} 

As a last setup step, we define the output.
\begin{lstlisting}[language=XML]	  
<meshOutput id="source_dpdt" inputFilterIds="pressure_time_derivative">
	<outputFile>
		<hdf5 extension="cfs" compressionLevel="6" externalFiles="no"/>
  <!-- compression level (greater number = bigger compression; default=1) -->
	</outputFile>
	<saveResults>
		<result resultName="acouRhsLoad">
			<allRegions/>
		</result>
	</saveResults>
</meshOutput>
\end{lstlisting} 
Finally, run the openCFS-Data interpolation inside the shell by the command \textit{cfsdat calc\_dpdt}.

\subsection{Acoustic simulation}
For the propagation simulation to obtain the acoustic field, we define the following openCFS XML-file.

The basic structure, including the XML header, looks the following:
\begin{lstlisting}[language=XML]	
<?xml version="1.0"?>
<cfsSimulation xmlns:xsi="http://www.w3.org/2001/XMLSchema-instance"
    xmlns="http://www.cfs++.org/simulation">
    
    FILE_FORMATS
    
	DOMAIN
    
    <sequenceStep>
        
		ANALYSIS_TYPE
        
        <pdeList>

			PDE
            
        </pdeList>
        
        SOLVER_SETTINGS
        
        
    </sequenceStep>
\end{lstlisting} 

\paragraph{Snippet: File formats}
\begin{lstlisting}[language=XML]	
<fileFormats>
	<input>
		<hdf5 fileName="source_dpdt.cfs"/>
	</input>
	<output>
		<hdf5 id="h5"/>
		<text id="txt"/>
	</output>
	<materialData file="mat.xml" format="xml"/>
</fileFormats>
\end{lstlisting}

\paragraph{Snippet: Computational domain}
We define the material for all regions as well as the [non-conforming interface](../../../../Tutorials/Features/ncinterfaces.md). Additionally, we define a microphone point at which we want to obtain the acoustic pressure for post-processing.
\begin{lstlisting}[language=XML]	
<domain geometryType="3d">
	<regionList>
		<region name="LARYNX" material="air"/>
		<region name="VT" material="air"/>
		<region name="PR" material="air"/>
		<region name="PML" material="air"/>
	</regionList>
	<ncInterfaceList>
		<ncInterface name="IF1" masterSide="IF_VT" slaveSide="IF_PR"/>
	</ncInterfaceList>
	<nodeList>
		<nodes name="mic">
			<coord x="0.24657" y="0.009" z="-0.049069"/>
		</nodes>
	</nodeList>
</domain>
\end{lstlisting} 

\paragraph{Snippet: Analysis Type}
We compute a transient simulation of one full cycle of the vocal folds' motion, which are oscillation at $f_0=148 \mathrm{Hz}$. Therefore, we define \textbf{numsteps}$= 675$ time steps with a step size of $\Delta t=1\cdot 10^{-5} s$.

\begin{lstlisting}[language=XML]	
<analysis>
	<transient>
		<numSteps>675</numSteps> 
		<deltaT>1e-5</deltaT>
	</transient>
</analysis>
\end{lstlisting} 

\paragraph{Snippet: Acoustic PDE}

Including definition of \textbf{PML regions}, \textbf{non-conforming interfaces} and \textbf{Temporal Blending} of the RHS source with the blending function $f(t)$
\begin{equation}
	f(t) = \left\{ \begin{array}{lll}
	-\frac{1}{\rho_{0} c^2}\left[1-cos\left(\frac{0.5 \pi}{8\cdot 10^{-4}} t\right)\right] & \mathrm{if} & t < 8\cdot 10^{-4} s\\
	-\frac{1}{\rho_{0} c^2} & \mathrm{else}
	\end{array}\right.
\end{equation}
with the density $\rho_{0} = 1.204 \frac{kg}{m^3}$ and the speed of sound $c = 343.4 \frac{m}{s}$. 
As output, we define the \textbf{acouPotentialD1} $\frac{\partial \psi_a}{\partial t }$, saved both for the whole domain as \textbf{.cfs} output and at the microphone point as \textbf{.txt} output.

\begin{lstlisting}[language=XML]	
<acoustic formulation="acouPotential" timeStepAlpha="-0.3">
    <regionList>
        <region name="LARYNX"/>
        <region name="VT"/>
        <region name="PR"/>
        <region name="PML" dampingId="dampPML2"/>
    </regionList>
    <ncInterfaceList>
        <ncInterface name="IF1" formulation="Nitsche" nitscheFactor="50"/>
    </ncInterfaceList>
    <dampingList>
        <pml id="dampPML2">
            <propRegion>
                <direction comp="x"     min="0.19"      max="0.275"/>
                <direction comp="y"     min="-0.051"    max="0.069"/>
                <direction comp="z"     min="-0.0525"   max="0.0675"/>
            </propRegion>
            <type>inverseDist</type>
            <dampFactor>1.0</dampFactor>
        </pml>
    </dampingList>
    <bcsAndLoads>
        <absorbingBCs volumeRegion="LARYNX" name="IF_ABC"/>
        <rhsValues name="LARYNX">
            <grid>
                <defaultGrid quantity="acouRhsLoad" dependtype="GENERAL" >
                    <globalFactor> ((t lt 8e-4)? 
                    (-1.0)/(1.204*343.4*343.4)*(1*(1-(cos(0.5*pi/8e-4*t))^2)) :
                    (-1.0)/(1.204*343.4*343.4))</globalFactor>
                </defaultGrid>
            </grid>
        </rhsValues>
        <rhsValues name="VT">
            <grid>
                <defaultGrid quantity="acouRhsLoad" dependtype="GENERAL">
                    <globalFactor> ((t lt 8e-4)? 
                    (-1.0)/(1.204*343.4*343.4)*(1*(1-(cos(0.5*pi/8e-4*t))^2)) : 
                    (-1.0)/(1.204*343.4*343.4))</globalFactor>
                </defaultGrid>
            </grid>
        </rhsValues>
    </bcsAndLoads>
    <storeResults>
        <nodeResult type="acouPotentialD1">
            <allRegions/>                   
            <nodeList>
                <nodes name="mic" outputIds="txt"/>
            </nodeList>           
        </nodeResult>                
    </storeResults>
</acoustic>
\end{lstlisting}

\paragraph{Snippet: Solver Settings}

\begin{lstlisting}[language=XML]	
<linearSystems>
	<system>
		<solverList>
			<pardiso id="default">
			</pardiso>
		</solverList>
	</system>
</linearSystems>
\end{lstlisting} 

Finally, run the simulation inside the shell using \textit{cfs propagation}.

\subsection{Results}
As a simulation result, we obtain the acoustic pressure field by multiplying the time derivative of the scalar acoustic potential with the density $p_a = \rho_0 \frac{\partial \psi_a}{\partial t }$,
which can be visualized with \textit{Paraview}
\begin{figure}[ht!]
    \centering
    \includegraphics[scale=0.2]{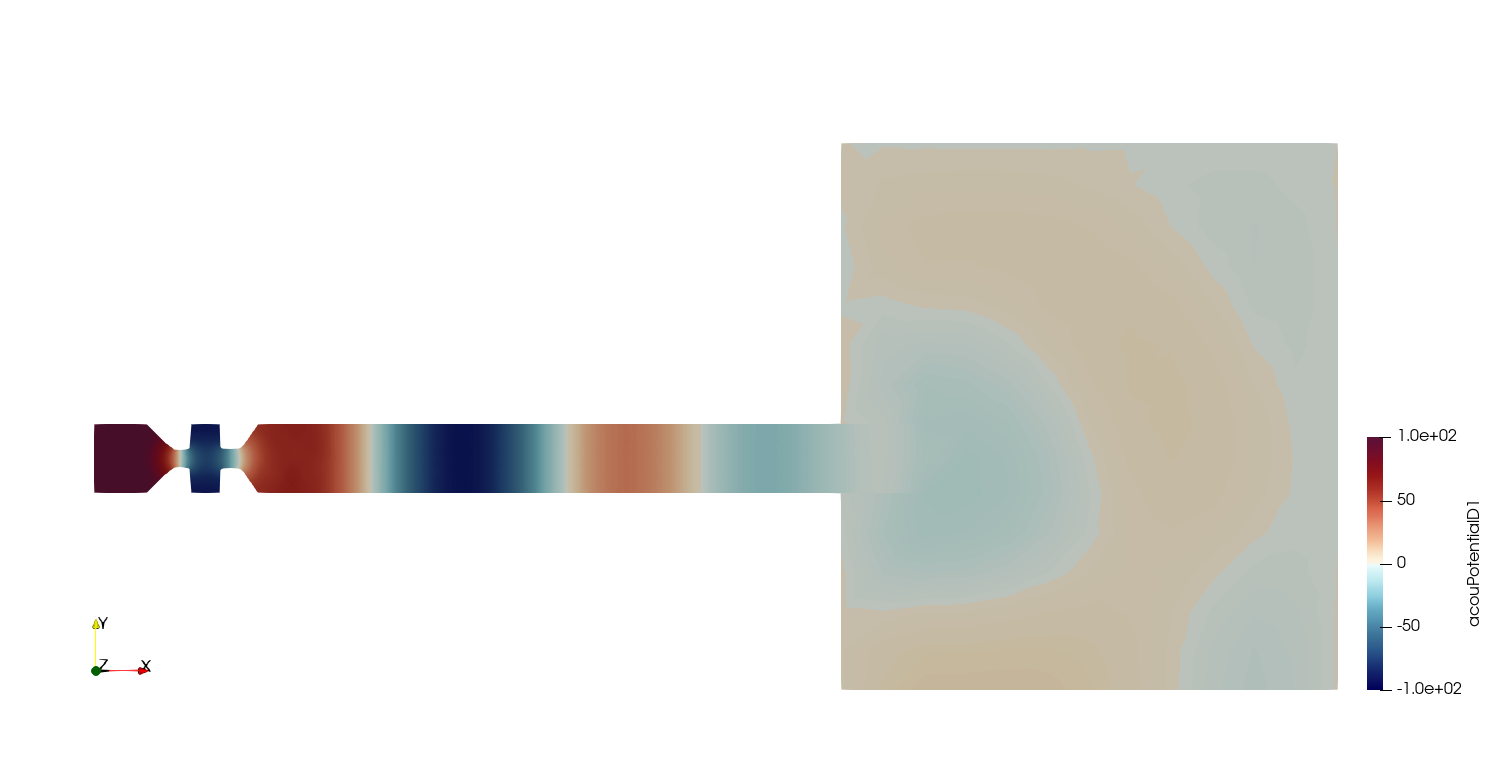}
    \caption{$\frac{\partial \psi_a}{\partial t }$ in the (x,y) plane.}
    \label{fig:my_label}
\end{figure}
\begin{figure}[ht!]
    \centering
    \includegraphics[scale=0.2]{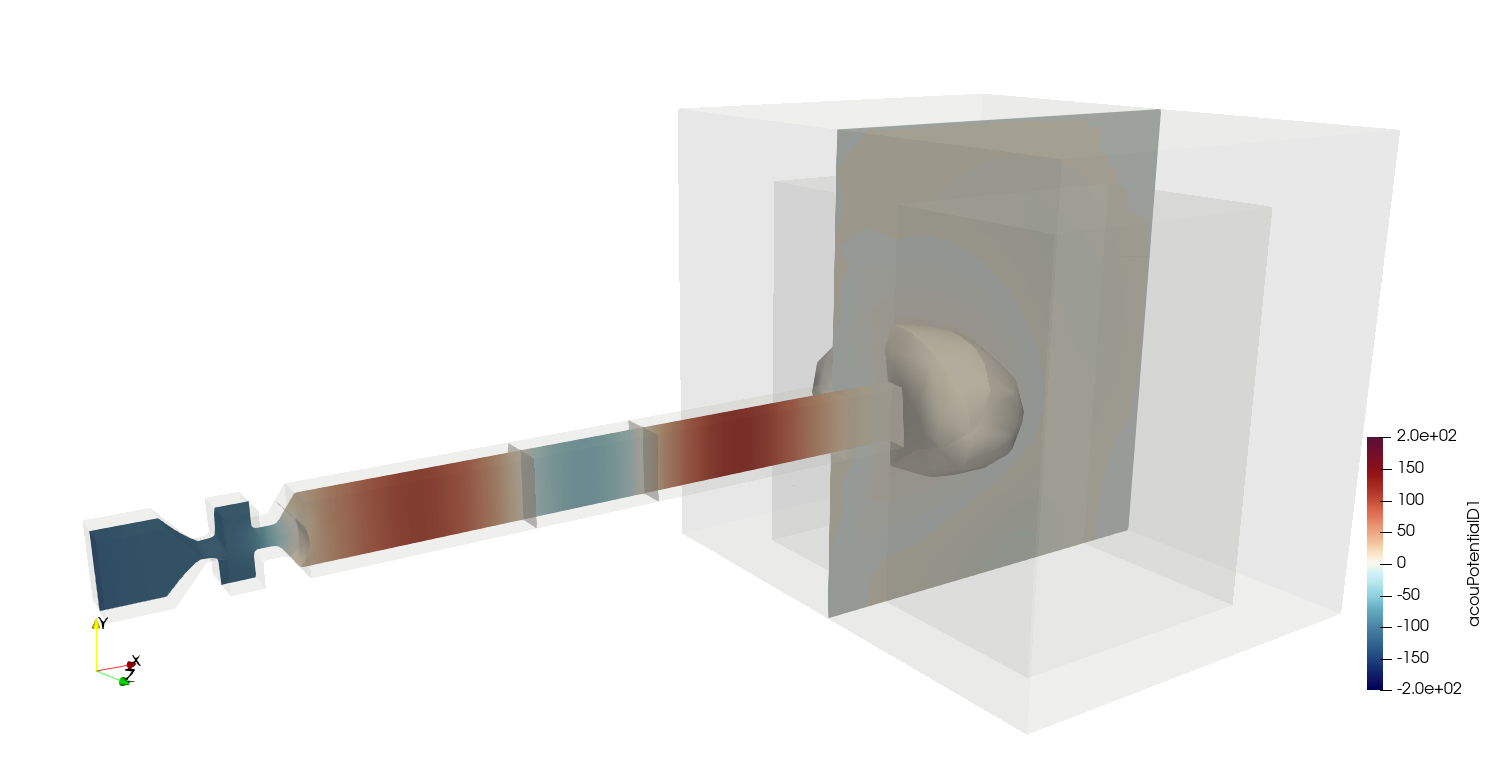}
    \caption{Contour planes at $\frac{\partial \psi_a}{\partial t } = 5 \frac{m^2}{s^2}$.}
    \label{fig:my_label2}
\end{figure}

Further interpretations can be done by calculating the Amplitude Spectral Density of the acoustic pressure at the microphone point and plot the spectral distribution of the sound pressure level (SPL). The following figure compares the simulation data of 20 vocal fold oscillation cycles with a measurement signal.
\begin{figure}[ht!]
    \centering
    \includegraphics[scale=0.2]{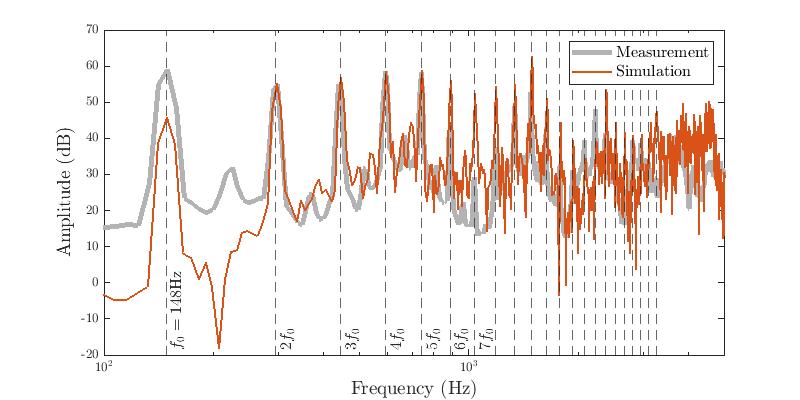}
    \caption{Amplitude Spectral Density of the acoustic pressure at a microphone point.}
    \label{fig:my_label3}
\end{figure}

\section{Conclusion}

Based on the presented information, the validation setup of the simVoice model can be reproduced using openCFS-DATA and openCFS. In conclusion, the interested reader can compute its own human phonation process using the hybrid aeroacoustic workflow. This simVoice model has high potential, and this training article should give an overview of how to work with such a model. Parts of the simulation data and further details can be provided upon request.

\section{Acknowledge}
We would like to acknowledge the authors of openCFS.

\bibliographystyle{abbrv}
\bibliography{references}  






\end{document}